\documentclass[prl,a4paper,twocolumn,superscriptaddress]{revtex4-1}

\usepackage[utf8x]{inputenc}
\usepackage{amssymb}
\usepackage{amsmath}
\usepackage{graphicx}
\usepackage{psfrag}
\usepackage{latexsym}
\usepackage{hyperref}

\newcommand{\be}{\begin{equation}}
\newcommand{\ee}{\end{equation}}
\newcommand{\bea}{\begin{eqnarray}}
\newcommand{\eea}{\end{eqnarray}}
\newcommand{\ba}{\begin{eqnarray}}
\newcommand{\ea}{\end{eqnarray}}

\newcommand{\beq}{\begin{equation}}
\newcommand{\eeq}{\end{equation}}
\newcommand{\beqa}{\begin{eqnarray}}
\newcommand{\eeqa}{\end{eqnarray}}
\newcommand{\beqar}{\begin{eqnarray*}}
\newcommand{\eeqar}{\end{eqnarray*}}

\newcommand{\eq}{\begin{equation}}
\newcommand{\eqx}{\end{equation}}
\newcommand{\eqn}{\begin{eqnarray}}
\newcommand{\eqnx}{\end{eqnarray}}

\newcommand{\reef}[1]{(\ref{#1})}

\newcommand{\eg}{{\it e.g.,}\ }
\newcommand{\ie}{{\it i.e.,}\ }

\newcommand{\rhov}{\rho_{\tiny V}}
\newcommand{\rhoB}{\rho_{\tiny B}}
\newcommand{\Hv}{H_{\tiny V}}
\newcommand{\bB}{{\overline B}}
\newcommand{\bV}{{\overline V}}

\begin{document}

\author{Jan de Boer}\email{j.deboer@uva.nl}

\affiliation{\it Institute for Theoretical Physics, University of Amsterdam,
1090 GL Amsterdam, The Netherlands} 

\author{Michal P. Heller}\email{mheller@perimeterinstitute.ca}

\altaffiliation[On leave from: ]{\emph{National Centre for Nuclear Research,  Ho{\.z}a 69, 00-681 Warsaw, Poland.}}

\affiliation{\it Perimeter Institute for Theoretical Physics, Waterloo, Ontario N2L 2Y5, Canada} 

\author{Robert C. Myers}\email{rmyers@perimeterinstitute.ca}

\affiliation{\it Perimeter Institute for Theoretical Physics, Waterloo, Ontario N2L 2Y5, Canada} 

\author{Yasha Neiman}\email{yashula@gmail.com}

\affiliation{\it Perimeter Institute for Theoretical Physics, Waterloo, Ontario N2L 2Y5, Canada} 

\title{Holographic de Sitter Geometry from Entanglement in Conformal Field Theory} 
%\title{Entanglement Holography} 
%\title{Dynamical fields in de Sitter from CFT entanglement} 

\begin{abstract}

We demonstrate that for general conformal field theories (CFTs), the entanglement for small perturbations of the vacuum is organized in a novel holographic way. %The first order change in the entanglement entropy, viewed as a function on the space of spherical entangling regions in a constant time slice, turns out to be 
For spherical entangling regions in a constant time slice, perturbations in the entanglement entropy are solutions of a Klein-Gordon equation in an auxiliary de Sitter (dS) spacetime. The role of the emergent time-like direction in dS is played by the size of the entangling sphere. For CFTs with extra conserved charges, \eg higher-spin charges, we show that each charge gives rise to a separate dynamical scalar field in dS.

%We demonstrate that in conformal field theories (CFT) the entanglement for small perturbations of the vacuum is organized in a novel holographic way. For spherical entangling regions in a constant time slice, the perturbation in the entanglement entropy is a solution of a Klein-Gordon equation in an auxiliary de Sitter (dS) spacetime. The role of the emergent time-like direction in dS is played by the size of the entangling sphere. For CFTs with extra conserved charges, \eg higher spin charges, we show that each charge gives rise to a separate dynamical scalar field in dS.

%Our results hold for any CFT in any number of dimensions and rely solely on the applicability of the first law of entanglement and the use of spherical entangling surfaces.

%We demonstrate that in conformal field theories (CFTs) the entanglement for small perturbations of the vacuum is organized in a novel holographic way in which the extra dimension is time-like. For spherical entangling regions in a constant time slice, the perturbation in the entanglement entropy is a solution of a Klein-Gordon equation in an auxiliary de Sitter (dS) spacetime. The role of the emergent time-like direction in dS is played by the scale (\ie radius) of the entangling sphere. Our results hold for any CFT in any number of dimensions and rely solely on the applicability of the first law of entanglement and the use of spherical entangling surfaces. For CFTs with extra conserved charges, \eg higher spin charges, we show that each charge gives rise to a separate dynamical field in dS.

\end{abstract}

\maketitle

\noindent \emph{Introduction and summary.--} Understanding the structure of quantum entanglement has emerged as a central question in elucidating novel emergent phenomena in complex quantum systems. This issue arises in a wide variety of research areas, ranging from condensed matter physics to quantum gravity. In the former, entanglement entropy distinguishes exotic phases of matter, such as quantum Hall fluids or spin liquids, while in quantum gravity, entanglement plays a key role in the emergence of quantum spacetime and also of the gravitational equations of motion \cite{mvr,eom1,eom2,eom3,eom4}. A great deal of recent progress in this area has come from the AdS/CFT correspondence. In this framework, entanglement entropies in the boundary CFT are encoded holographically in terms of the Bekenstein-Hawking entropy of extremal surfaces in the dual bulk spacetime \cite{rt1,rt2}.

%In this paper, we show that the entanglement structure of any CFT has a novel holographic description in terms of an auxiliary de Sitter (dS) geometry in one higher dimension. As is typical, scale in the CFT emerges as the additional holographic direction, but in the present construction, it plays the role of {\it time} in the dS geometry (with a standard Lorentzian signature). Our finding is reminiscent of the apparent causal structure inherent to the Multiscale Entanglement Renormalization Ansatz (MERA) \cite{MERA} and a recent proposal regarding the its holographic interpretation \cite{bartek2}. We emphasize though that our construction is not related to the standard AdS/CFT correspondence and applies for any CFT in any number of dimensions, without any requirement of a large central charge or strong coupling. 

In this paper, we show that the entanglement structure of any CFT has a novel holographic description in terms of an auxiliary de Sitter (dS) geometry. As is typical, scale in the CFT emerges as the extra holographic direction, but in the present construction the scale plays the role of {\it time} in the dS geometry. This structure is reminiscent of the effective causal structure inherent to the Multiscale Entanglement Renormalization Ansatz \cite{MERA,beny} and the interpretation of these tensor networks within AdS$_{3}$/CFT$_{2}$ correspondence \cite{bartek2}. Quite remarkably, we find that the emergent dS geometry is the arena for a local dynamics, in which one of the degrees of freedom is identified with the perturbations of the entanglement entropy (EE). We emphasize though that our construction is not related to the standard AdS/CFT correspondence and applies for any CFT in any number of dimensions, without any requirement of a large central charge or strong coupling. 

We begin with a $d$-dimensional CFT in its vacuum state in flat spacetime. Now consider evaluating the EE for a ($d-1$)-dimensional ball $B$ of radius $R$ centered at $\vec{x}$ on a fixed time slice. As we review below, for weakly excited states,
the change in the EE is fixed by the expectation value of the energy density:
\be
\label{eq.deltaS}
\delta S(B) = 2 \pi \int_{B}\!\! d^{d-1} x' \ \frac{R^{2} - |\vec{x} - \vec{x}'|^{2}}{2 R} \,\langle T_{t t}(\vec{x}')\rangle \,.
\ee
While this result is now fairly well known, it went completely unnoticed that the integration kernel in the above expression is a boundary-to-bulk propagator in $d$-dimensional dS geometry where the radius $R$ plays the role of the \emph{time-like} coordinate
\be
\label{eq.deSitter}
ds^{2} =  \frac{L^{2}}{R^{2}} \left( - dR^{2} + d\vec{x}^{2} \right)\,.
\ee
Hence, $\delta S(\vec{x},R)$ obeys the Klein-Gordon equation 
\be
\label{eq.wave}
\left(\nabla_{a} \nabla^{a}%\, \delta S 
- m^{2} \right)\, \delta S = 0\,,
\ee
in this auxiliary dS, where the mass is given by
\be
\label{eq.mass}
m^{2} L^{2} = -d.
\ee
This result is the focal point of the Letter. Further, as we demonstrate below, for CFTs with extra global (\eg higher-spin) charges, there exists one additional dynamical field in dS for each charge.

With Eq.~\reef{eq.wave}, the asymptotic boundary data (\ie the behavior at $R = 0$) is the expectation value of the energy density $\langle T_{t t}\rangle$, which sets $\delta S$ at very small scales.  Then the EE perturbations  at larger scales are determined by the local Lorentzian propagation into the dS geometry. Hence the EE for small excitations of the vacuum state in \emph{any} CFT is organized with respect to scale in a novel Lorentzian holographic manner. As we discuss below, the choice of the asymptotic boundary data implicit in Eq.~\reef{eq.deltaS} is precisely that needed to remove the unstable modes associated with the mass term~\eqref{eq.mass} being tachyonic.

We would like to point out that in the AdS/CFT framework, the same wave equation \reef{eq.wave} implicitly appears in \cite{Nozaki:2013vta,Bhattacharya:2013bna} but their derivation relied on the use of holographic EE and the Einstein equations in the bulk. However, we reiterate that our construction is not connected to the AdS/CFT correspondence. As we discuss below, the present holographic propagation of $\delta S$ relies solely on the so-called `first law of entanglement.' 

%We note that in the AdS/CFT framework, the same wave equation \reef{eq.wave} implicitly appears in \cite{Nozaki:2013vta,Bhattacharya:2013bna} but their derivation relied on the use of holographic EE and the Einstein equations in the bulk. However, we reiterate that our construction is not connected to the AdS/CFT correspondence. As we discuss below, the present holographic propagation of $\delta S$ relies solely on the so-called `first law of entanglement.' 

Finally, for pure states, the EE of any ball matches that of its complement. This condition imposes an antipodal symmetry on the solutions of Eq.~\reef{eq.wave}. Combining this property with Eq.~\reef{eq.deltaS}, we find novel constraints on the profile of the energy density --- see Eqs.~\reef{eq.moments_flat} and~\reef{eq.moments_sphere}.

%Finally, let us re-emphasize that our construction applies for any CFT in any number of dimensions, without any requirement of a large central charge or strong coupling and, as such, it is not directly related to the standard holography. Notably, Refs.~\cite{Nozaki:2013vta,Bhattacharya:2013bna} arrived implicitly at Eq.~\eqref{eq.wave} by considering holographic EE and Einstein's equations with negative cosmological constant.

\vspace{6 pt}

\noindent \emph{First Law as Lorentzian Propagation.--} To evaluate EE in a quantum field theory, we divide a constant time slice into two parts, a region $V$ and its complement $\bV$. Upon tracing out the degrees of freedom in $\bV$, we are left with the reduced density matrix $\rhov$ describing the remaining degrees of freedom in the region $V$. The EE 
is then evaluated with the standard expression for the von Neumann entropy:
\be
S(V)=-\textrm{Tr}\left(\rhov \log\rhov\right)\,.
\label{eq.Sent}
\ee
Since the reduced density matrix is both hermitian and positive
semidefinite, it can be expressed as
 \be
\rhov=\frac{e^{-\Hv}}{\mathrm{tr} \, e^{-\Hv}} 
\label{eq.Hmod},
 \ee
where the hermitian operator $\Hv$ is known as the modular Hamiltonian \cite{Haag:1992hx}.
%The modular Hamiltonian (\ref{eq.Hmod}) governs the change in the entanglement entropy for states being 
Now for a small perturbation of the reduced density matrix $\rhov + \delta \rho$, one can show that the change in the entanglement entropy (\ref{eq.Sent}) is given by~\cite{relE}
\be
\label{eq.1stlaw}
\delta S = \delta \langle \Hv \rangle,
\ee
where $\delta \langle H \rangle$ denotes the change in the expectation value of the modular Hamiltonian associated with the original density matrix~$\rhov$. 
%Further, we have used $\mathrm{Tr}  \, \delta \rho = 0$ and terms of ${\cal O} \left( \delta \rho^{2} \right)$ have been neglected here. 
%The above relation can be also deduced from the properties of the relative entropy, which provides an useful measure of distance between states. 
Eq.~\eqref{eq.1stlaw} is commonly called the first law of entanglement, as it is a quantum analog of the first law of thermodynamics. 
%The latter corresponds to taking $V$ to be the whole space and $\rho$ being the thermal density matrix. In this case the entanglement Hamiltonian is given by the microscopic Hamiltonian divided by the temperature $T$.

The cases in which the entanglement Hamiltonian is local are rare. One special case is for a CFT in its vacuum state in $d$-dimensional Minkowski spacetime $\mathbb{R}^{1,d-1}$ and where the region of interest is a spherical ball $B$. 
In this case, the entanglement Hamiltonian takes the simple form
\be
\label{eq.HCFT}
H_B = 2 \pi \int_{B}\!\! d^{d-1} x' \ \frac{R^{2} - |\vec{x} - \vec{x}'|^{2}}{2 R}\, T_{t t}(\vec{x}') \,,
\ee
where the integral is taken over the ball centered at position $\vec{x}$ and with radius $R$, and $T_{t t}$ is the energy density operator. Now, combining this expression with the first law \eqref{eq.1stlaw}, we find that for spherical regions, the change in the entanglement entropy from the vacuum to weakly excited states is given by Eq.~\reef{eq.deltaS}.
Implicitly, we are using the fact that the expectation value of $T_{t t}$ vanishes in the vacuum.

As noted above, the integration kernel in Eq.~\eqref{eq.deltaS} is a boundary-to-bulk propagator in $d$-dimensional dS space with the metric \reef{eq.deSitter}. 
%, where $R$ plays the role of the \emph{time-like} coordinate. 
Now, one easily verifies that the perturbation $\delta S$ obeys the wave equation \reef{eq.wave} on this auxiliary dS. In general, this equation has two independent asymptotic solutions 
which to leading order take the form
\be
\label{eq.bound}
\delta S \ \stackrel{R\to0}{=}\  F(\vec{x})/R + f(\vec{x})\,R^d + \cdots \,.
\ee
Hence, our wave equation admits boundary data with conformal weights $\Delta = -1$ and $d$, corresponding to $F(\vec{x})$ and $f(\vec{x})$, respectively. We can then identify Eq.~\eqref{eq.deltaS} as the solution with  
\be
\label{eq.bound2}
F(\vec{x})=0 \quad{\rm and} \quad
f(\vec{x})=\frac{\pi^{\frac{d+1}2} }{\Gamma\left(\frac{d+3}2\right)}\,\langle T_{t t}(\vec{x})\rangle\,.
\ee
Therefore, the expectation value of the energy density sets $\delta S$ at very small scales (\ie $R\to0$).  Then the EE perturbations  at larger scales are determined by the Lorentzian propagation into the dS geometry, according to Eq.~\reef{eq.wave}. 
Thus, the EE for excited states around the vacuum is organized with respect to scale in a novel Lorentzian holographic manner. Again, this result applies for any CFT in any number of dimensions $d$ and relies solely on the applicability of the first law. 

%xyz
We also reiterate that the above choice of boundary data \eqref{eq.bound2} precisely removes `non-normalizable' or unstable modes associated with the tachyonic mass term~\eqref{eq.mass}.

At this point, let us note that the wave equation \reef{eq.wave} is a covariant expression and so it can be rewritten in terms of any coordinate system on the dS geometry. As usual, changing coordinates in the bulk amounts to choosing a new conformal frame in the boundary CFT. Hence our holographic construction readily extends to the CFT in any conformally flat background. The cylindrical background $\mathbb{R}\times \mathbb{S}^{d-1}$ is of particular interest below. In this case, a constant time slice corresponds a round $(d-1)$-sphere and the corresponding wave equation then appears in global coordinates on the dS space, \eg
$ds^2=L^2(-d\tau^2 + \cosh(\tau)^2 d\Omega_{d-1})$. Explicit examples of propagation in this cylindrical conformal frame and in the flat frame are given in the supplemental material, which includes also Refs.~\cite{PO,Herzog:2014fra}. 
%rcm
There we also give an alternative derivation of the wave equation \eqref{eq.wave} based on group-theoretic arguments.%qwe

\vspace{6 pt}

\begin{figure}
\includegraphics[width=0.45\textwidth]{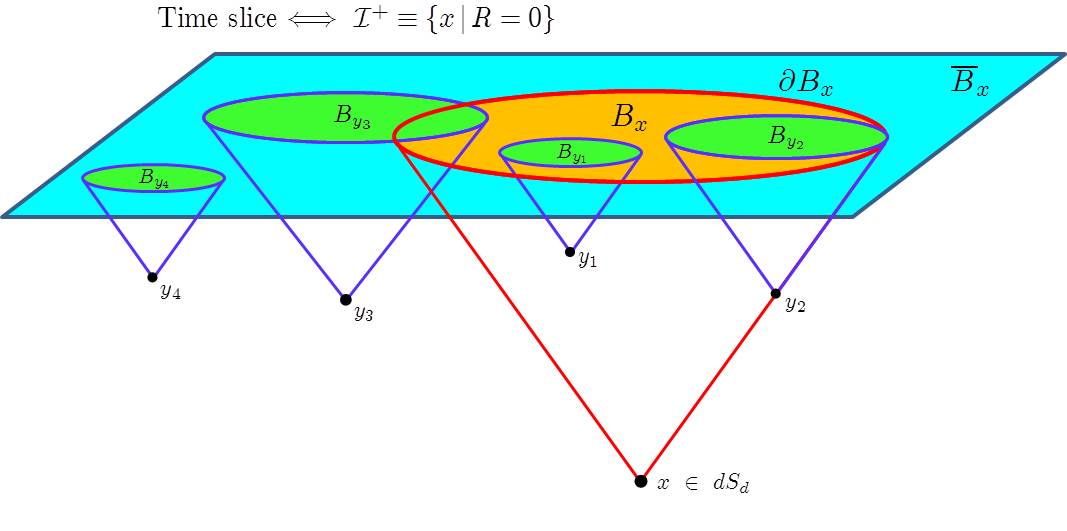}
\caption{One-to-one mapping between points in dS geometry and balls on its future asymptotic boundary $\mathcal{I}^+$. We identify the latter with the constant time slice in a given CFT. The future lightcone of the bulk point intersects $\mathcal{I}^+$ on the boundary of the corresponding ball.} 
\label{fig.dSmap}
\end{figure}

\noindent \emph{Auxiliary de Sitter Geometry.--} The relation between balls on a constant time slice of a $d$-dimensional CFT and the dS geometry is easily inferred as follows: Implicitly, we have identified our time slice with the future~\footnote{We could have just as well chosen the past asymptotic boundary.} asymptotic boundary $\mathcal{I}^+$ of dS. Now, for any bulk point $x\in$ dS, the intersection of the inside of the future lightcone of $x$ with $\mathcal{I}^+$ is then a ball-shaped region, see Fig.~\ref{fig.dSmap}. This establishes a one-to-one map between points in dS and balls on the time slice.

\begin{figure}
\includegraphics[height=0.25\textheight]{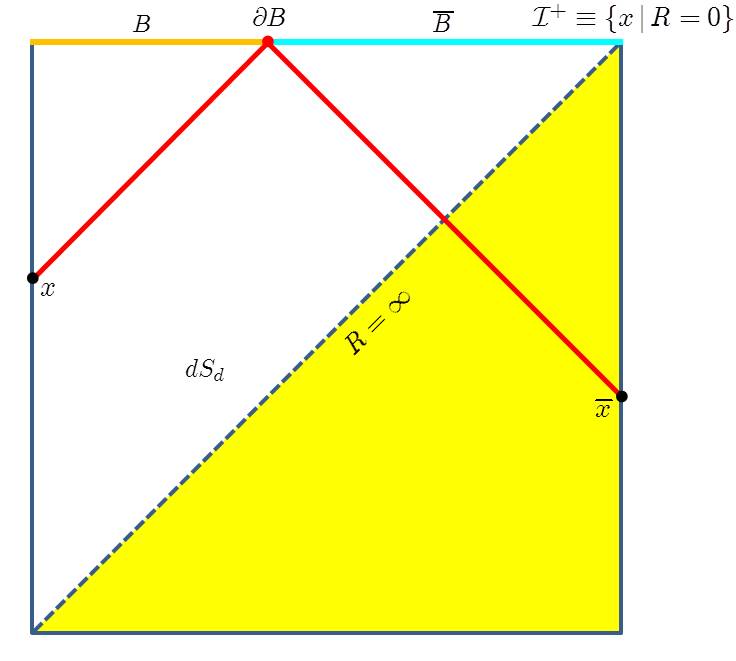}
\caption{Penrose diagram of dS. Points correspond to $\mathbb{S}^{d-2}$, whereas horizontal lines represent $\mathbb{S}^{d-1}$. $\partial B$ represents the spherical entangling surface and $B$ (orange) and $\bar B$ (cyan) its interior and exterior. Red lines represent corresponding lightcones and $x$ and $\bar{x}$ their tips, \ie bulk points in~dS corresponding to $B$ and $\bar{B}$ regions. Note that $\bar x$~is~the~antipode~of~$x$.} 
\label{fig.dSelliptic}
\end{figure}

In much of our discussion, the time slice has the topology of $\mathbb{R}^{d-1}$, as assumed in the modular Hamiltonian \eqref{eq.HCFT}. Then this map, and the dS metric in Eq.~\reef{eq.deSitter}, only covers half of the dS geometry, \ie the expanding Poincare patch. The missing half can be identified with the \emph{complementary}  exterior regions $\bB$ on this flat slice. This is more easily seen by going to a conformal frame where our time slice has $\mathbb{S}^{d-1}$ topology. As described above, this corresponds to choosing global coordinates in the bulk dS space. Each spherical entangling surface then defines two ball-shaped regions covering complementary domains on the $\mathbb{S}^{d-1}$. These two balls are identified with antipodal points in the dS geometry, see Fig.~\ref{fig.dSelliptic}. 

Furthermore, the Lorentzian structure of the dS geometry can be given a geometric interpretation directly in terms of balls on the CFT time slice. Causal relationships between points $x\in$ dS become topological relationships between the corresponding balls $B_{x}$, as illustrated in Fig.~\ref{fig.dSmap}. In particular, we have: $1$) $y_1$ is in the time-like future of $x$  $\Leftrightarrow$  $B_{y_1}$ is contained within $B_x$; $2$) $y_2$ is in the null future of $x$ $\Leftrightarrow$ $B_{y_2}$ lies within $B_x$ but $\partial B_{y_2}$ is tangent to $\partial B_x$ at one point; and $3$) $x$ and $y_3$ (or $y_4$) are space-like separated $\Leftrightarrow$ the domain of $B_{y_3}$ (or $B_{y_4}$) extends beyond $B_x$ and vice versa. The latter can be refined in a number of ways by taking into account the relationship with the antipodal point $\overline{x}$, \eg $3a)$ $x$ and $y_4$ are space-like separated and $y_4$ is in the time-like future of $\overline{x}$  $\Leftrightarrow$  $B_{y_4}$ is completely outside of $B_x$. With $d=2$, this structure reduces to the relations discussed  in Ref.~\cite{bartek2} for one-dimensional intervals. %an arbitrary number of dimensions.
% or  $3b$) $x$ and $y$ are space-like separated and $x$ is in the null future of $\overline{y}$ $\Leftrightarrow$ $B_x$ lies outside of $B_y$ but $\partial B_x$ is tangent to $\partial B_y$ from the outside.
% $3c$) is in the time-like past of $\overline{y}$  $\Leftrightarrow$ $B_x$ and $B_y$ intersect on an annular region with $\partial B_x$ entirely within $\partial B_y$; $3d$) $x$ is in the null past of $\overline{y}$ $\Leftrightarrow$ $B_x$ and $B_y$ intersect on an annular region which degenerates at one point where $\partial B_x$ is tangent to $\partial B_y$.

What remains is to give meaning to the scale factor of the dS metric in terms of the CFT time slice. This can be fixed by considering two concentric balls in $\mathbb{R}^{d-1}$, with radii $R$ and $R + dR$. We then define the `time-like distance' between these balls as $L \, dR/R$, where the constant $L$ becomes the dS radius. This is the only definition that respects conformal symmetry on the time slice.

Finally, we note that the fact that we can associate Lorentzian order between balls on a constant time slice in Minkowski spacetime does not imply the existence of local dynamics respecting this structure. Hence the appearance of the wave equation \eqref{eq.wave} is rather remarkable.

%\section{Antipodal Symmetry}

\vspace{6 pt}

\noindent \emph{Antipodal Symmetry.--} If we consider excitations which are globally \emph{pure} states, they must satisfy the constraint that the EE inside each ball must equal that in its complement. Hence
\be
\label{eq.equal_entropies}
\delta S(B) = \delta S(\bB)
\ee
In the holographic dS picture, this corresponds to \emph{antipodally even} configurations $\delta S(x)$, \ie $\delta S(x) = \delta S(\bar x)$. 
%Essentially, the propagation for this class of excitations is restricted to elliptic dS space, i.e. the quotient of dS under the antipodal map \cite{Schrodinger:2010zz,Folacci:1986gr}. 

Considering general solutions of the wave equation \eqref{eq.wave}, this antipodal symmetry imposes a relation between the two classes of boundary data in Eq.~\reef{eq.bound} of the form:
\be
\label{eq.antipodal_symmetry}
\frac{2\pi^{\frac{d-1}2} }{\Gamma\left(\frac{d+3}2\right)}\ F(\vec x) = \int d^{d-1} x' \ |\vec{x} - \vec{x}'|^{2}\ f(\vec{x}') \,,
\ee
where the integral is over the future boundary $\mathcal{I}^+$ of dS. 
Now, recall that our entropy configurations \eqref{eq.deltaS} are not general solutions, but ones with vanishing $\Delta = -1$ data.  Hence combining Eq.~\reef{eq.bound2} with the antipodal symmetry constraint \eqref{eq.antipodal_symmetry}, we arrive at a constraint on the energy density profile:
%\be \label{eq.energy_constraint_raw}
$\int d^{d-1} x' \, |\vec{x} - \vec{x}'|^{2} \, \langle T_{tt}(\vec x')\rangle = 0$.
% \, . \ee
For Eq.~\reef{eq.equal_entropies} to hold for all balls, this constraint %\eqref{eq.energy_constraint_raw} 
must be satisfied for all $\vec x$. This is equivalent to the vanishing of the following moments of $\langle T_{tt}(\vec x)\rangle$:
\begin{align}
 \begin{split}
   \int d^{d-1}x&\, \langle T_{tt}(\vec x)\rangle = 0 \, , \quad \int d^{d-1}x\ \vec x\ \langle T_{tt}(\vec x)\rangle = 0\, , \\
%   \int d^{d-1}x\ \vec x\ \langle T_{tt}(\vec x)\rangle &= 0 \, , \\ 
 &\,\,\,\mathrm{and} \quad  \int d^{d-1}x\  \vec x^2\,  \langle T_{tt}(\vec x)\rangle  = 0 \, .   
 \end{split} \label{eq.moments_flat}
\end{align}
Note that these constraints can be identified as the vanishing of the expectation value of the total energy, the boost generators, and the temporal generator of special conformal transformations, respectively. Focusing on the vanishing of the total energy, we must recall that we are working at leading order in a small perturbation above the vacuum. This means that the energy difference will appear at higher orders in the expansion. At this point, we note that while $S(B)=S(\bB)$ for all $B$ would certainly indicate that the underlying state is globally pure, some mixed states will still satisfy the leading order constraint \reef{eq.equal_entropies} in our perturbative construction.

We note that the constraints \eqref{eq.moments_flat} can also be derived directly in the CFT, by comparing Eq.~\eqref{eq.deltaS} to the analogous expression for the entropy in the ball's complement:
\be
\delta S(\bB) = 2 \pi \int_{\bB}\!\! d^{d-1} x' \ \frac{|\vec{x} - \vec{x}'|^{2} - R^2}{2 R} \,\langle T_{t t}(\vec{x}')\rangle \,.
\ee
%rcm
Let us mention as well that the constraints \eqref{eq.moments_flat} %and Eq.~\eqref{eq.moments_sphere} could have been 
can also be derived from the results of Ref.~\cite{Blanco:2013lea}.
Finally, we include the analogue of Eq.~\eqref{eq.moments_flat} for the conformal frame where the time slice is spherical:
\begin{align}
 \int d^{d-1}n\ \langle T_{tt}(n)\rangle = 0 \, , \ \int d^{d-1}n\ n^I\, \langle T_{tt}(n)\rangle = 0 \, . \label{eq.moments_sphere}
\end{align}
Here, points on $\mathbb{S}^{d-1}$ are parametrized by $d$-dimensional unit space-like vectors $n^I$, with volume form $d^{d-1}n$.

\vspace{6 pt}

\noindent \emph{Extension to Higher-Spin Charges.--} In this section, we discuss a generalization of our holographic dS construction, based on viewing the stress tensor as the special spin-2 case of a conserved symmetric traceless current $T_{\mu_1\dots\mu_s}$ with arbitrary spin $s\geq 1$. The $s=1$ case is an ordinary charge current $J_\mu$. The $s>2$ case is relevant for CFTs with higher-spin symmetry. These include free theories in all dimensions, as well as some non-trivial theories in $d=2$ (\eg see \cite{Gaber,Maldacena:2011jn} and references therein). 
\begin{figure}
\includegraphics[height=0.25\textheight]{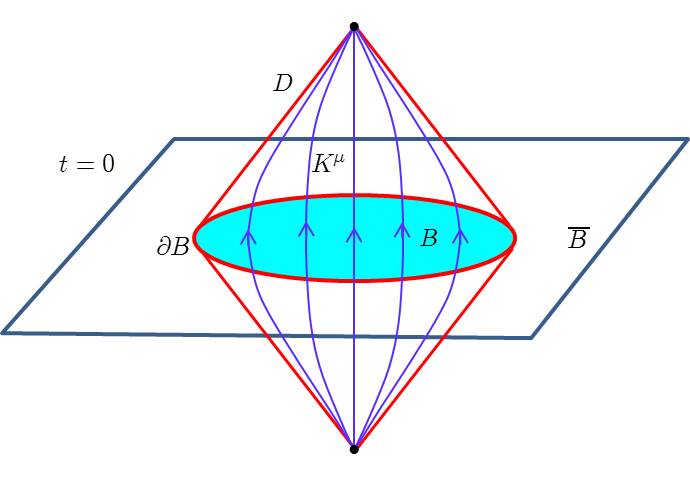}
\caption{The domain of dependence $D$ of a ball $B$ enclosed by a given spherical entangling surface. Blue lines are tangent to the conformal Killing vector field $K^{\mu}$, which gives rise to conserved charges \eqref{eq.HS}.} 
\label{fig.ballcausdev}
\end{figure}

First, let us note that the expression \eqref{eq.HCFT} for the modular Hamiltonian has a covariant meaning in the CFT background spacetime. It is the flux through~$B$ of the conserved current $J^{(2)}_\mu\equiv T_{\mu\nu} K^\nu$ where $K^\mu$ is the (time-like) conformal Killing vector that preserves the boundary of $B$, see Fig.~\ref{fig.ballcausdev}. This suggests a natural generalization to the spin-$s$ case: for each  $T_{\mu_1\dots\mu_s}$, we define a charge $Q^{(s)}$ as the flux through $B$ of the current $J^{(s)}_\mu\equiv T_{\mu\nu_2\dots\nu_s} K^{\nu_2}\dots K^{\nu_s}$. This charge is given by the following integral on our chosen time slice
\be
\label{eq.HS}
Q^{(s)} = (2\pi)^{s-1} \int_{B}\!\! d^{d-1} x' \left(\frac{R^{2} - |\vec{x} - \vec{x}'|^{2}}{2 R}\right)^{s-1} T_{tt\dots t}(\vec{x}') \,.
\ee
Note that all of these currents are conserved, \ie $\nabla^\mu J^{(s)}_\mu=0$, and hence evaluating this flux through any hypersurface which is homologous to $B$ yields the same charge $Q^{(s)}$. The modular Hamiltonian \eqref{eq.HCFT} is the special case $s=2$, \ie $H_B=Q^{(2)}$. One can use the new charges to construct new reduced density matrices $\rhoB \sim\exp\left[-\sum \mu_s Q^{(s)}\right]$ to refine the measurement of the entanglement between $B$ and $\overline{B}$. Ref.~\cite{alex} studied this approach for $s=1$, where $Q^{(1)}$ is an ordinary charge. Higher-spin charges $Q^{(s)}$ with $s>2$ were discussed in Ref.~\cite{Hijano:2014sqa} for two-dimensional CFTs. 

We now observe that the integration kernel in Eq.~\eqref{eq.HS} is again a boundary-to-bulk propagator in dS, for a scalar with mass given by:
\be
\label{eq.mass_general}
m^2 L^2 = -(s-1)(d+s-2) \,.
\ee
Thus, the charges \eqref{eq.HS} can all be interpreted as scalar fields in dS, obeying a Lorentzian wave equation with mass as in Eq.~\eqref{eq.mass_general}. For an ordinary charge with $s=1$, the bulk field in dS is massless, while charges with $s\geq 2$ correspond to a discrete series of tachyonic masses. Of course, these mass values are precisely those that allow boundary data with conformal weights $d+s-2$, in agreement with the weights of the densities $T_{tt\dots t}$. 

The antipodal symmetry constraints \eqref{eq.moments_flat} generalize most cleanly when recast in their $\mathbb{S}^{d-1}$ form \eqref{eq.moments_sphere}. Then the analogous constraint for general spin $s$ is the vanishing of the first $s$ moments of $T_{tt\dots t}(n^I)$. For even (odd) $s$, these constraints lead to antipodally even (odd) wave solutions in dS, so that the charge $Q^{(s)}$ in every ball equals plus (minus) the analogous charge in its complement. For example, in the case of $s=1$, this reduces to the vanishing of the total charge on the time slice.

\vspace{6 pt}

\noindent \emph{Outlook.--} We demonstrated that the conformal group of a $d$-dimensional CFT induces a Lorentzian geometry on the space of balls on a constant time slice, which corresponds to~dS$_d$. Remarkably, the entanglement of small excitations of the CFT vacuum is governed a local wave equation~\reef{eq.wave} on this auxiliary geometry. The expectation value $\langle T_{00}\rangle$ is the asymptotic boundary data in the dS space, fixing $\delta S$ at small scales. The EE perturbations  at larger scales are then determined by the propagation into the dS geometry, according to Eq.~\reef{eq.wave}. Hence the EE in any CFT is organized with respect to scale in a novel Lorentzian holographic manner. We also gave the generalization for CFTs with extra global (\eg higher spin) charges, with one dynamical field in dS for each charge.

Our holographic propagation of $\delta S$ relies solely on the first law of entanglement for spherical regions. The first law must apply not just for a particular sphere but for all spheres (and their complements) on a given time slice. For pure states, the EE of any ball matches that of its complement and this introduces an antipodal symmetry on the solutions in the dS space. Combining this property \reef{eq.equal_entropies} with the first law, we found novel constraints on the profile of the energy density, \ie Eqs.~\reef{eq.moments_flat} and \reef{eq.moments_sphere}.

Looking forward, it remains to be seen if our holographic construction can be extended to provide a full description of the CFT in terms a local theory of interacting fields, including the metric, propagating in the dS spacetime. Such a theory would then provide a novel example of the dS/CFT correspondence (\eg see \cite{andy8,edward,Balasubramanian:2002zh}) in which the boundary CFT is {\it unitary}.

%. The present construction suggests a new perspective on the the dS/CFT correspondence, since the boundary CFT is unitary.
%our local bulk fields are associated with nonlocal observables in the boundary unitary CFT.
% \comment{???} 

%Progress in this direction may come from a number of related questions. The first of these is elucidating how the scale of the dS geometry is fixed. We would expect that $L$ will be determined in terms of CFT data through an understanding of the dynamics of the holographic geometry. A related issue would be understanding how EE perturbations around excited states (rather than around the vacuum) of the CFT might be described in terms of a holographic geometry. Similarly, it would be important to develop an extension of our framework for CFTs perturbed by a relevant operator. 

The present construction is closely related to the proposed description of EE in two-dimensional CFTs in terms of integral geometry \cite{bartek2}. Hence, integral geometry may provide an interesting perspective to further extend our holographic construction. It may also be that our new construction will provide useful insight into extending the proposal of Ref.~\cite{bartek2} to higher dimensions.

%We should add that the solution for the latter questions has been proposed for $d=2$, where the dual geometry is determined by the full EE on intervals in the boundary state \cite{bartek2}.

%A further question would be whether there is a physical role for the weight $-1$ boundary data $F(\vec{x})$ appearing in Eq.~\reef{eq.bound}. We might note that since the dS field has a `tachyonic' mass \reef{eq.mass} --- see also Eq.~\reef{eq.mass_general} --- it should be unstable but we can interpret setting $F(\vec{x})=0$ as  fine tuning the runaway modes to zero.  

Of course, a full holographic description would require understanding the time dependence of quantities in the CFT. The natural starting point here would be to consider spherical regions not simply on a fixed time slice but throughout the $d$-dimensional spacetime of the CFT. The group-theoretic construction presented in the supplemental material suggests that $\delta S$ now obeys a wave equation on the coset $SO(2,d)/[SO(1,d-1)\times SO(1,1)]$, which interestingly has multiple time directions. We are currently studying this framework in further detail.\\%qwe

%The group-theoretic construction, encapsulated by Eqs.~\eqref{eq.generator} and \eqref{eq.casdeltaS}, suggests that $\delta S$ now obeys a wave equation on the coset
%$SO(2,d)/[SO(1,d-1)\times SO(1,1)]$, which interestingly has multiple time directions. We are currently studying this framework in further detail.\\

%seems well-suited to address this point.
%The intuition developed here leads to the curious suggestion of a holographic description with two time-like directions.\\
% We are currently studying this fascinating question using appropriate generalizations of the group-theoretic arguments from Eq.~\eqref{eq.generator}.

%In this framework, the group theoretic argument suggests that action of the full conformal group $SO(2,d)$ on balls in the entire spacetime and the stabilizer of a single sphere becomes $SO(1,d-1)\times SO(1,1)$. Hence, the perturbations $\delta S$ will obey a wave equation of the form \reef{eq.casdeltaS} but now on the coset
%$SO(2,d)/[SO(1,d-1)\times SO(1,1)]$. \comment{do we still get $\nabla^2T_{tt}=-d\,T_{tt}$??} Interestingly, in this case, the auxiliary space is 2$d$-dimensional and has two time-like directions. We are currently studying this framework in further detail.\\
%\newpage

\vspace{6 pt}

\begin{acknowledgements}
We would like to thank H.~Casini, B.~Czech, V.~Hubeny, M.~Johnson, L.~Lamprou, A.~Lewkowycz, J.~Maldacena, S.~McCandlish, J.~Sully, and especially P.~Caputa, J.~Jottar, M.~Rangamani and S.~Ross for valuable comments and correspondence. We are also grateful to G.~Vidal for multiple discussions and collaboration on related subjects. Research at Perimeter Institute is supported by the Government of Canada through
Industry Canada and by the Province of Ontario through the Ministry of
Research \& Innovation. RCM and YN also acknowledge support from NSERC Discovery grants. RCM acknowledges funding from the Canadian Institute for Advanced Research. JB is supported in part by the research programme of
the Foundation for Fundamental Research on Matter (FOM), which is part of the Netherlands Organization for Scientific Research (NWO).

\end{acknowledgements}

\section*{Supplemental material}

\noindent \emph{Explicit Examples of States.--} Local propagation of $\delta S$ in auxiliary de Sitter geometry holds for any states which are small perturbations of the vacuum state everywhere in space. In particular, we require that for the excitations under consideration, the first law applies not just for a particular sphere but for all spheres (and their complements) on a given time slice.  Below we discuss two examples of such states.

\begin{figure}
\includegraphics[width=0.305\textwidth]{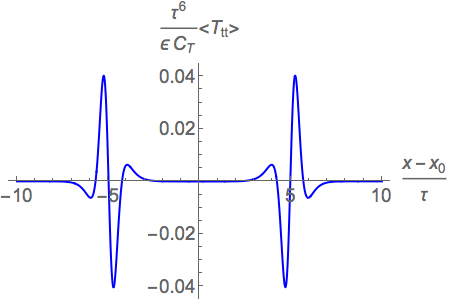}\\
\vspace{6 pt}
\includegraphics[width=0.305\textwidth]{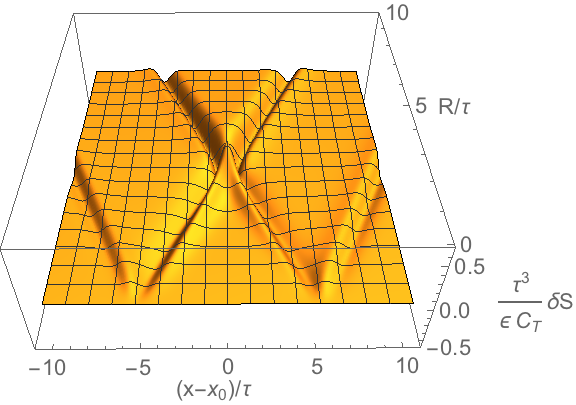}
\caption{Top: Rescaled energy density $\langle T_{t t}\rangle$ for the state (\ref{eq.phi}) with $d=3$ plotted along the $x$-axis at $t = t_{0} + 5\, \tau$, see Eq.~\eqref{eq.f1}. % For $t>t_{0}$ the energy is a sum of identical left- and right-moving contributions, which are given by the original energy density rescaled by a factor of $2$ and appropriately translated.\\
Bottom: Corresponding change in the entanglement entropy $\delta S$ as a function of dS time $R$ and position along the $x$-axis. The causal nature of  propagation is clearly visible.
% for the state (\ref{eq.phi}) at $t = t_{0} + 5 \, \tau$. The causal nature of  propagation is clearly visible. One also sees that the boundary data consists of $2$ localized bumps of energy separated by the distance of $10 \, \tau$.
} 
\label{fig.dSex}
\end{figure}

Let us first consider a pure state in a $d$-dimensional CFT on a plane, which is created by an infinitesimal insertion of the energy density operator $T_{t t}$ at time $t_{0} + i \, \tau$ and position $\vec{x}_{0}$
\be
\label{eq.phi}
| \phi \rangle = \left(1 + \epsilon \, T_{t t}\right) |0 \rangle\,,
\ee
see also Ref.~[19]. The parameter $\epsilon$ is taken to be small in the sense of $\epsilon / \tau^{d} \ll 1$. The evolution in imaginary time $\tau$ is included to regulate potential UV divergences and ensures that the state is a small perturbation of the vacuum. The energy density of the state \eqref{eq.phi} is determined by the two-point function of the stress tensor \cite{PO} 
\bea
&&\langle \phi | T_{t t} (t,x) | \phi \rangle =  \epsilon \, C_T\, 
\Bigg[ \frac{1}{(\Delta x^2 -(\Delta t+i \, \tau)^2 )^{d}} \label{eq.f1}\\&&\qquad\quad\times \Bigg(
\frac{(\Delta x^2 +(\Delta t+i \, \tau)^2 )^{2}}{(\Delta x^2 -(\Delta t+i \, \tau)^2 )^{2}}-\frac1d\Bigg) + \mathrm{c.c.} \Bigg]+ {\cal O}(\epsilon^{2})\, ,
\nonumber
% \epsilon \langle 0 | T_{t t} (t_{0} - i \eta, x_{0}) \, T_{t t} (t,x) | 0 \rangle \nonumber \\
%&&  + \epsilon \langle 0 | T_{t t} (t,x) \, T_{t t} (t_{0} + i \eta, x_{0}) | 0 \rangle
\eea
where $C_T$ is the central charge and we have defined $\Delta x^2=|\vec{x}-\vec{x}_0|^2$ and $\Delta t^2=|t-t_0|^2$. An explicit illustration of the energy density for this state and corresponding dS propagation of the perturbation in the EE is shown in Fig.~\ref{fig.dSex}. Note that the energy density~\eqref{eq.f1} is a spherical shell expanding out from $(t_0,\vec{x}_0)$ at the speed of light.  As expected from our general argument, the energy density profile \eqref{eq.f1} obeys the constraints (15) and, hence, the holographic propagation respects the antipodal symmetry on the auxiliary dS background. 

Our second example is the following mixed state %density matrix
\be
\label{eq.etaE}
\rho = |0\rangle \langle 0| + \eta \, | E \rangle \langle E |,
\ee
where $|E\rangle$ is an energy eigenstate (with constant energy density), and $\eta$ is a small parameter. In this case, we assume that the constant time slice has topology $\mathbb{S}^{d-1}$ with radius $r$. Let us now look at ($d$--2)-dimensional spherical entangling surfaces surrounding a cap of the $\mathbb{S}^{d-1}$ specified by the angle $\theta_{0}$. The first law reads now
\bea\label{gosh}
\delta S = && 2 \pi \int_{0}^{\theta_{0}}  r^{d-1}\Omega_{d-2}\, \sin^{d-2}\!{\theta}\ d\theta \times  \\ && \qquad \times\ r \, \frac{\cos\theta - \cos\theta_{0}}{\sin{\theta_{0}}}\ \times\ \frac{\eta\ E}{r^{d-1}\Omega_{d-1}} \,,\nonumber
\eea
where $\Omega_n=2\pi^{(n+1)/2}/\Gamma\left(\frac{n+1}2\right)$ denotes the volume of a unit $\mathbb{S}^{n}$. The factors in the integrand then correspond to, in order, the volume element of $\mathbb{S}^{d-1}$, the boundary-to-bulk propagator for dS in global coordinates, and the (constant) expectation value of the energy density. A special case of this expression appears in Ref.~\cite{Herzog:2014fra}, which discusses universal thermal corrections to the vacuum entanglement entropy. There, the energy is given by
$E = \frac{\Delta}{r}$, where $\Delta$ is the smallest scaling dimension in the spectrum (apart from the identity), % of operators including the energy-momentum tensor and all primaries not equal to the identity, 
and $\eta$ is the product of the degeneracy of the energy eigenstate and the corresponding Boltzmann factor, \ie  $\eta = g \, e^{-\beta \Delta/r}$.
%In the formula above, $g$ denotes possible degeneracy among the lowest lying operators and $\beta$ is the inverse temperature. 
Clearly, this and other mixed states of the %schematic 
form \eqref{eq.etaE} violate the first constraint in Eq.~\eqref{eq.moments_sphere}. Hence, the corresponding $\delta S$ propagates on dS without antipodal symmetry.

Note that $\delta S$ in Eq.~\reef{gosh} diverges as $\theta_0\to\pi$, \ie as the dS propagation reaches the past boundary ${\cal I}^-$. This divergence is related to
a breakdown of the first law and corresponding free propagation in dS space when
$\sin\theta_0\sim\eta E r$ (with $\theta_0>\pi/2$).

\noindent \emph{Alternative derivation of wave equation on dS geometry.--} Let us now present another perspective on the wave equation (3). The conformal group relevant for a $d$-dimensional CFT is $SO(2,d)$. However, only the subgroup $SO(1,d)$ leaves a constant time slice invariant. Hence the corresponding spherical entangling surfaces are mapped onto one another under the action of $SO(1,d)$. Now considering the perturbations $\delta S$ for these ball-shaped regions, the $SO(1,d)$ generators $K_i$ act as 
\be
\label{eq.generator}
\partial_{K_{i}} \delta S\left[\langle T_{t t} \rangle\right] =- \delta S\left[ \langle \partial_{K_{i}} T_{t t} \rangle \right]
\,.
\ee
Here, the $\partial_{K_{i}}$ on the RHS can be viewed as generating an ``active'' conformal transformation that changes the CFT state, while the $\partial_{K_{i}}$ on the LHS generates a ``passive'' transformation that instead changes the spherical entangling surface.
Comparing now the ``active'' and ``passive'' action of the quadratic Casimir of $SO(1,d)$, \mbox{$\nabla^{2} \equiv c_{i j} \partial_{K_{i}} \partial_{K_{j}}$}, we get
\be
\label{eq.casdeltaS}
\nabla^{2}  \delta S\left[\langle T_{t t} \rangle\right] = \delta S \left[\langle \nabla^{2} T_{t t} \rangle\right]
=-d\,\delta S\left[\langle T_{t t} \rangle\right] \,,
\ee
where the second expression above uses the linearity in $T_{t t}$ of the modular Hamiltonian \eqref{eq.HCFT}. 
Further, the last expression appears after using the fact that the energy density transforms as a scalar of weight $d$ with respect to the $SO(1,d)$ subgroup. Now, a particular spherical entangling surface is left invariant by the stabilizer group
 $SO(1,d-1)$. Hence, on the LHS of Eq.~\reef{eq.casdeltaS}, the nontrivial action of $\nabla^{2}$ is on the coset space \mbox{$SO(1,d)/SO(1,d-1)$}. The latter coset is precisely the anticipated $d$-dimensional dS geometry, and $\nabla^{2}$ becomes the d'Alembertian on this space. Hence this group theoretic approach produces precisely the Klein-Gordon equation (3) on the auxiliary dS space. Note that this analysis implicitly normalizes the dS radius $L$ to unity. 
 
Finally, let us mention that the group theoretic argument above can be also generalized to the higher-spin case and, as expected, yields the mass given by Eq.~\eqref{eq.mass_general}.

\bibliography{dSEE3}{}

\end{document}